\documentclass[aps,twocolumn]{revtex4}
\usepackage[utf8]{inputenc}
\usepackage{amsmath,amssymb}
\usepackage{array}
\usepackage{graphicx}
\usepackage{color}                                           
\begin{document}

\title{Landauer's Principle as a Criterion for Thermodynamic Consistency in Generalized Black Hole Entropies}
\author{Fatemeh Sadeghi\footnote{fatemeh.sadeghi96@ph.iut.ac.ir}}
\address{Department of Physics, Isfahan University of Technology, Isfahan 84156-83111, Iran\\
Department of Physics, College of
Science, Shiraz University, Shiraz 71454, Iran}
\author{Ahmad Sheykhi\footnote{asheykhi@shirazu.ac.ir}}
\address{Department of Physics, College of
Science, Shiraz University, Shiraz 71454, Iran\\
Biruni Observatory, College of Science, Shiraz University, Shiraz
71454, Iran}

\begin{abstract}
\noindent Under the assumption that black hole horizon area is
quantized, each Hawking evaporation step, during which the black
hole loses mass and transitions to a lower area level, is
interpreted as the erasure of one bit of information. In this
paper, by employing Landauer's principle, we test the consistency
of various black hole entropy relations with this
information-theoretic framework. For the Bekenstein-Hawking
entropy, the energy emitted per step saturates the Landauer bound.
We extend this analysis to a broad class of generalized entropy
models, yielding three distinct outcomes. In the first category,
Landauer's principle imposes constraints on the Hawking
temperature and, consequently, on the black hole mass. In the
second, it restricts the free parameters of the entropy model. The
third category, exemplified by Kaniadakis entropy, proves
incompatible with Landauer's principle. For all compatible models,
we derive the area quantization parameter and the corresponding
area spectrum. While this parameter is constant for
Bekenstein-Hawking entropy, it becomes level-number-dependent for
many generalized models. Nonetheless, the relative spacing between
successive area levels vanishes in the classical limit. Our
findings point to a deep link between information theory and black
hole thermodynamics.

\end{abstract}
\maketitle

\newpage
\section{\label{Intro}Introduction}
In 1970, Christodoulou studied reversible transformations for the
Kerr black hole. He showed that a particle can enter the black
hole without changing the horizon area. This happens when the
particle carries the minimum energy required for capture, has the
appropriate angular momentum, and its radial velocity vanishes at
the event horizon. Although the black hole mass and angular
momentum change, the horizon area remains unchanged. He called
such processes reversible transformations
\cite{christodoulou1970reversible}. Christodoulou and Ruffini
extended this analysis to the Kerr-Newman black hole in 1971
\cite{christodoulou1971reversible}. They included electric charge
and generalized the concept of the irreducible mass
\cite{christodoulou1971reversible}. Bekenstein later applied the
same idea to the Reissner-Nordström black hole. He showed that
the horizon area also remains unchanged when a charged particle is
absorbed under the corresponding reversible conditions
\cite{bekenstein1973black}. These results led Bekenstein to
propose that the event horizon area is a classical adiabatic
invariant. He then used Ehrenfest's adiabatic principle. This
principle states that every classical adiabatic invariant
corresponds to an observable with a discrete spectrum in quantum
theory. Based on this argument, Bekenstein proposed that the black
hole horizon area should have a discrete eigenvalue spectrum
\cite{bekenstein1973black}. He developed this idea within the old
quantum theory by applying the Bohr-Sommerfeld quantization rule.
This work provided the first physical motivation for black hole
horizon area quantization
\cite{bekenstein1973black,bekenstein1974generalized}.

This proposal raised another question. If the event horizon area
has a discrete spectrum, what is the statistical meaning of the
Bekenstein-Hawking area law of black hole entropy, namely
$S_{BH}=A/(4l_P^2)$? Mukhanov answered this question in 1986 by
proposing that each area level corresponds to a large number of
microscopic quantum states rather than a single state. The black
hole entropy is the logarithm of this degeneracy
\cite{mukhanov1986black}. The number of microstates at the $n$th
area level is $g_n=e^{S_n/k_B}$. This interpretation became the
basis of black hole statistical mechanics and thermodynamics
\cite{mukhanov1986black,wald2001thermodynamics,frolov2012black}.
Bekenstein and Mukhanov later proposed that the horizon area is
uniformly quantized, $A_n=\gamma\,l_P^2\,n$, where $\gamma$ is a
dimensionless constant and $n$ is the area quantum number
\cite{bekenstein1995spectroscopy}. For a Schwarzschild black hole,
the horizon area depends only on the black hole mass. Therefore, a
discrete area spectrum also implies a discrete mass spectrum. The
emitted energy cannot take arbitrary values. It must equal the
energy difference between two neighboring levels. Hawking
radiation can then be viewed as a sequence of quantum transitions
between discrete area levels \cite{bekenstein1995spectroscopy}.
These transitions occur only at discrete frequencies. However,
each area level has a large number of microscopic states. As a
result, the envelope of the emission lines remains approximately
thermal and agrees with the semiclassical Hawking spectrum
\cite{bekenstein1995spectroscopy}. The same picture also describes
energy absorption. When a black hole absorbs energy, it moves to a
higher area level. Hawking radiation corresponds to transitions to
lower levels. Every downward transition decreases the horizon area
and the Bekenstein-Hawking entropy. This picture raises another
question. Can the entropy decrease be interpreted in terms of
information theory? Can a quantum transition between two area
levels represent the loss or erasure of information? If the answer
is yes, Landauer's principle may provide a direct connection
between information theory and the discrete spectrum of the event
horizon. We begin by reviewing the relation between information
and thermodynamics.

The relationship between information and thermodynamics began with
Maxwell's demon. Szilard later gave this idea a quantitative form
by introducing the Szilard engine. In this model, measuring the
position of a particle allows work to be extracted from a thermal
reservoir. This result raises a simple question. How does the
second law of thermodynamics remain valid? Several decades later,
John von Neumann asked whether a computer operating at temperature
$T$ must always dissipate heat during computation
\cite{parrondo2015thermodynamics}. Landauer answered this question
and showed that heat dissipation is not a necessary feature of
every computational process, while it occurs only during logically
irreversible operations, most notably the erasure of one bit of
information \cite{landauer1961irreversibility}. This necessarily
requires the dissipation of at least $Q_{\min}=k_B\,T\,\ln2$ of
heat into the environment \cite{landauer1961irreversibility}.
Finally, Bennett completed this picture and discussed that the
irreversible step in Maxwell's demon is not the measurement
itself. It is the erasure of the memory that stores the
measurement outcome. When this thermodynamic cost is included, the
entropy increase of the environment compensates for the entropy
decrease of the system. The second law of thermodynamics therefore
remains valid \cite{bennett1982thermodynamics}.

Landauer's principle has been tested experimentally in a wide
range of physical systems. These experiments established it as a
fundamental principle of information thermodynamics
\cite{berut2012experimental,jun2014high}. Researchers have since
extended and reformulated Landauer's principle in both classical
and quantum information thermodynamics \cite{bennett2003notes,
maroney2009generalizing, lorenzo2015landauer,
chattopadhyay2025landauer}. In recent years, Landauer's principle
has attracted considerable attention in black hole physics. Recent
studies have examined information crossing the event horizon, the
generalized second law of black hole thermodynamics, the
connection between information erasure and Hawking evaporation,
and geometric interpretations of gravitational entropy related to
the Bekenstein-Hawking entropy \cite{fuchs1992landauer,
song2008information, kim2010black, cortes2024hawking,
isidro2026landauer}. Landauer's principle has also been applied to
the Bekenstein-Hawking entropy and to a few generalized entropy
models. These studies derived the corresponding area quantization
parameters and area spectra \cite{ananias2026generalized}.

Although the Bekenstein-Hawking entropy successfully describes
black hole thermodynamics, many approaches to quantum gravity
predict that this relation changes when quantum gravitational
effects become important. Similar modifications can also arise
from the microscopic structure of spacetime, the fractal nature of
the event horizon, or long range statistical correlations. These
ideas have motivated many generalized entropy formalisms over the
past decades. Some models originate from quantum gravity
corrections. Others are based on generalized statistical
mechanics. Some modify the geometry of the event horizon or
introduce phenomenological corrections to the Bekenstein-Hawking
area law \cite{ashtekar2000, kaniadakis2002statistical,
kaniadakis2006towards, zhang2008black, banerjee2008quantum1,
banerjee2008quantum, banerjee2008noncommutative,
sheykhi2010thermodynamics, sheykhi2011power, tsallis2013black,
czinner2016renyi, majhi2017non, jahromi2018generalized,
sheykhi2018modified, barrow2020area, abreu2021black, liu2022non,
sheykhi2025mond,Goh1,Goh2,Luci1,Luci2,sheykhi2025emergence}.
Generalized entropy models now play an important role in black
hole thermodynamics, cosmology, quantum gravity, and information
theory. This work examines whether these entropy models satisfy
Landauer's principle. This analysis may help clarify the
connection between information theory and black hole
thermodynamics.

Our work differs from \cite{ananias2026generalized}, where they
derived the area quantization parameter for the Bekenstein-Hawking
entropy and a few modified entropy models. They also showed that
the Bekenstein-Hawking entropy saturates the Landauer bound under
the assumption that each Hawking emission corresponds to the
erasure of one bit of information \cite{ananias2026generalized}.
However, they did not systematically assess the consistency of
generalized entropy models with Landauer's principle. One example
is the R\'enyi entropy \cite{czinner2016renyi}. The authors of
\cite{ananias2026generalized} assumed a negative entropy parameter
to avoid a divergence of the area quantization parameter. However,
it did not examine whether this choice is compatible with
Landauer's principle. In the present work, we adopt Landauer's
principle as the primary criterion for testing the consistency of
generalized entropy models. We first impose the Landauer
inequality on the emitted energy. We then derive the corresponding
consistency constraints for each entropy model. For the R\'enyi
entropy, our analysis requires a positive entropy parameter in
order to satisfy Landauer's principle. The area quantization
analysis further restricts this parameter to values smaller than
unity. For the Kaniadakis entropy
\cite{kaniadakis2002statistical,kaniadakis2006towards}, our
analysis shows that it is incompatible with the Landauer
criterion. After establishing these constraints, we derive the
corresponding area quantization parameters and area spectra for
all entropy models that satisfy Landauer's principle. We summarize
the main results in a unified table and a comparative figure. The
table and figure allow a direct comparison of the generalized
entropy models.

The paper is organized as follows. In Sec. \ref{sec2}, we review
the Bekenstein-Mukhanov approach to black hole area quantization
and compares it with the Landauer based approach. We also discuss
the similarities and differences between the two frameworks. In
Sec. \ref{sec3}, we apply Landauer's principle to a class of
generalized entropy formalisms for which it leads to constraints
on the ambient temperature. In Sec. \ref{sec4}, we study entropy
formalisms in which Landauer's principle constrains the free
entropy parameters. In Sec. \ref{sec5}, we examine the
compatibility of the Kaniadakis entropy with Landauer's principle
and shows that this entropy formalism is incompatible with the
principle. We summarize the Landauer constraints in a unified
table and presents the corresponding area quantization parameters
in a comparative figure in Sec. \ref{sec6}. The last section is
devoted to conclusions and outlook for future investigations.
\section{Event Horizon Area Quantization: The Bekenstein-Mukhanov and Landauer Approaches}\label{sec2}
In this section, we review the main results of the
Bekenstein-Mukhanov approach \cite{bekenstein1995spectroscopy} and
the Landauer based interpretation \cite{ananias2026generalized}.
Our goal is to establish the theoretical framework required for
the subsequent sections. These results form the basis of the
analysis of generalized entropy formalisms presented later in this
work. We first review the Bekenstein-Mukhanov approach and derive
the spacing between consecutive event horizon area levels. We then
derive the corresponding area level spacing within the framework
of Landauer's principle. We conclude this section by comparing the
results of the two approaches. In the Bekenstein-Mukhanov
approach, the black hole event horizon area is assumed to be
uniformly quantized as \cite{bekenstein1995spectroscopy}
\begin{equation}\label{bm1}
A_n=\gamma\,l_P^2\,n,\qquad n=1,\,2,\,3,\,\ldots,
\end{equation}
where $\gamma$ is a dimensionless constant and
$l_P=\sqrt{\left(\hbar\,G\right)/c^3}$ is the Planck length.
They assumed that the number of microstates associated with the $n$th area level is
\begin{equation}\label{bm2}
W_n=k^n,
\end{equation}
where $k$ is a fixed integer greater than unity that represents
the number of microstates associated with each area quantum. Using
Eq. \eqref{bm2} in the Boltzmann-Gibbs entropy gives
\begin{equation}\label{bm3}
S_{BG}=k_B\,\ln\,W_n=k_B\,n\,\ln\,k.
\end{equation}
The Bekenstein-Hawking entropy at the area level $n$ is defined by
\begin{equation}\label{bm4}
S_{BH}=\frac{k_B\,A_n}{4\,l_P^2}.
\end{equation}
By equating Eqs. \eqref{bm3} and \eqref{bm4}, we obtain the event horizon area spectrum
\begin{equation}\label{bm5}
A_n=4\,l_P^2\,n\,\ln\,k.
\end{equation}
We compare Eqs. \eqref{bm1} and \eqref{bm5} and obtain the area quantization constant
\cite{bekenstein1995spectroscopy}
\begin{equation}\label{bm6}
\gamma=4\,\ln\,k.
\end{equation}
We now derive the same quantity within the framework of Landauer's
principle. We assume that each transition between two consecutive
area levels corresponds to the erasure of one bit of information.
According to Landauer's principle, the entropy change for the
erasure of one bit of information is
\cite{landauer1961irreversibility}
\begin{equation}\label{bm7}
\Delta\,S=k_B\,\ln\,2.
\end{equation}
Next, Eqs. \eqref{bm1} and \eqref{bm4} give the entropy difference between two consecutive area levels
\begin{equation}\label{bm8}
S_{n+1}-S_n=\frac{k_B\,\gamma}{4}.
\end{equation}
If each transition between two consecutive area levels corresponds
to the erasure of one bit of information, the entropy change
predicted by Landauer's principle equals the entropy difference
between two successive area levels. We equate Eqs. \eqref{bm7} and
\eqref{bm8} and obtain the area quantization constant
\cite{ananias2026generalized}
\begin{equation}\label{bm9}
\gamma=4\,\ln\,2.
\end{equation}
A comparison of Eqs. \eqref{bm6} and \eqref{bm9} shows that the
consistency of the Bekenstein-Mukhanov approach with the Landauer
based framework requires
\begin{equation}\label{bm10}
k=2.
\end{equation}
In this case, both approaches predict the same value for the
spacing constant between two consecutive area levels, namely
$\gamma=4\,\ln\,2$. This agreement shows that if each quantum
transition between two consecutive event horizon area levels
corresponds to the erasure of one bit of information, the
Bekenstein-Mukhanov model and Landauer's principle are fully
consistent with each other.

We next examine whether the discrete area spectrum approaches
continuous behavior in the classical limit, where
$n\rightarrow\infty$. This result should not be restricted to the
particular choice $k=2$. Therefore, the following analysis uses
the general expression for the area quantization parameter,
$\gamma=4\ln k$. We substitute this expression into Eq.
\eqref{bm1}. The area spectrum becomes
\begin{equation}\label{bma1}
A_n=4\,n\,l_P^2\,\ln\,k.
\end{equation}
The area difference between two consecutive levels is
\begin{equation}\label{bma2}
\Delta\,A_n=A_{\left(n+1\right)}-A_n=4\,l_P^2\,\ln\,k.
\end{equation}
Although Eq. \eqref{bma2} shows that the spacing between two
consecutive area levels is independent of the level number and
remains constant, the classical behavior of the spectrum depends
on the ratio of this spacing to the area itself. Eq. \eqref{bma1}
gives this ratio
\begin{equation}\label{bma3}
\frac{\Delta\,A_n}{A_n}=\frac{1}{n}.
\end{equation}
In the classical limit, we obtain
\begin{equation}\label{bma4}
\lim_{n\rightarrow\infty}\,\frac{\Delta\,A_n}{A_n}=0.
\end{equation}
Eq. \eqref{bma4} shows that although the absolute spacing between
two consecutive area levels remains constant, its ratio to the
horizon area approaches zero in the limit $n\rightarrow\infty$. At
macroscopic scales, the discrete area spectrum effectively
exhibits continuous behavior and correctly recovers the classical
limit.

Next, we show that the mass change associated with Hawking
evaporation of a Schwarzschild black hole also saturates the
Landauer bound. The event horizon area of a Schwarzschild black
hole is given by
\begin{equation}\label{as1}
A=4\,\pi\,r_s^2,
\end{equation}
where
\begin{equation}\label{as2}
r_s=\frac{2\,G\,M}{c^2},
\end{equation}
is the Schwarzschild radius. Substituting this relation into the
Bekenstein-Hawking entropy, Eq. \eqref{bm4}, and adopting $
\hbar=c=1 $, we obtain
\begin{equation}\label{bm11}
S_{BH}=4\,\pi\,G\,k_B\,M^2.
\end{equation}
Differentiating Eq. \eqref{bm11} with respect to the black hole mass, we obtain
\begin{equation}\label{bm12}
\frac{d\,S_{BH}}{d\,M}=8\,\pi\,G\,k_B\,M.
\end{equation}
For an infinitesimal mass variation, we have
\begin{equation}\label{bm13}
\Delta\,S_{BH}=8\,\pi\,G\,k_B\,M\,\Delta\,M.
\end{equation}
We now assume that each step of Hawking evaporation corresponds to
the erasure of one bit of information. According to Landauer's
principle, the minimum entropy change is
\begin{equation}\label{as3}
\Delta\,S_{BH}=k_B\,\ln\,2.
\end{equation}
We substitute this expression into Eq. \eqref{bm13} and obtain the
corresponding mass change for each evaporation step
\begin{equation}\label{bm14}
\Delta\,M=\frac{\ln\,2}{8\,\pi\,G\,M}.
\end{equation}
The Hawking temperature of a Schwarzschild black hole is
\begin{equation}\label{as4}
T=\frac{1}{8\,\pi\,G\,k_B\,M}.
\end{equation}
Eq. \eqref{bm14} can be written as
\cite{ananias2026generalized}
\begin{equation}\label{bm15}
\Delta\,M=k_B\,T\,\ln\,2.
\end{equation}
Eq. \eqref{bm15} shows that if each Hawking evaporation step
corresponds to the erasure of one bit of information, the emitted
energy reaches the minimum energy required by Landauer's
principle. Therefore, the Landauer bound is saturated.

The following sections use the framework reviewed in this section
to analyze generalized entropy formalisms. In this more general
setting, Landauer's principle takes the form
\cite{landauer1961irreversibility}
\begin{equation}\label{as5}
\Delta\,M\ge k_B\,T\,\ln\,2.
\end{equation}
We investigate the constraints imposed by this inequality on
either the Hawking temperature or the free parameters of each
entropy formalism. For the Bekenstein-Hawking entropy, this
inequality reduces to an equality. Therefore, the Landauer bound
is exactly saturated.
\section{Generalized Entropies Subject to Temperature Constraints Imposed by Landauer's Principle}\label{sec3}
In this section, we use Landauer's principle as a thermodynamic
criterion to examine generalized entropy formalisms. We consider
entropy models for which Landauer's principle directly constrains
the Hawking temperature. We derive these constraints for each
entropy model and discuss their physical implications.
\subsection{Mass-to-Horizon Entropy}
The Mass-to-Horizon entropy was introduced to establish
thermodynamic consistency between the Clausius relation, the
Hawking temperature, and generalized entropy formalisms. It
replaces the conventional linear relation between the black hole
mass and the horizon radius with a generalized relation. This
modification leads to the Mass-to-Horizon entropy
\cite{Goh1,Goh2}. This entropy reproduces the Bekenstein-Hawking
entropy in the appropriate limit. It also provides a unified
framework for a family of generalized entropy formalisms. The
Tsallis-Cirto \cite{tsallis2013black, sheykhi2018modified} and
Barrow \cite{barrow2020area} entropies appear as special cases for
suitable choices of its parameters. The Mass-to-Horizon entropy is
\cite{Luci2}
\begin{equation}\label{mh1}
S_{MH}=\frac{2\,\eta\,\lambda}{\eta+1}\,r_s^{\eta-1}\,S_{BH},
\end{equation}
where $ \lambda $ is a positive constant with dimensions $
[L]^{1-\eta} $, $ \eta $ is a non-negative real parameter, $r_s$
is the Schwarzschild horizon radius, and $ S_{BH} $ denotes the
Bekenstein-Hawking entropy. Using Eqs. \eqref{bm4} and
\eqref{as1}, Eq. \eqref{mh1} can be written as
\begin{equation}\label{mh1a}
S_{MH}=\frac{2\,\pi\,k_B\,\eta\,\lambda}{l_P^2\,\left(\eta+1\right)}\,r_s^{\eta+1},
\end{equation}
which reduces to the Bekenstein-Hawking entropy for $ \lambda=\eta=1 $. The Tsallis-Cirto entropy is
\cite{tsallis2013black}
\begin{equation}\label{t1}
S_{TC}=k_B\,\delta\,A^{\zeta},
\end{equation}
where $A=4\,\pi\,r_s^2$ is the event horizon area, and $\zeta$ and
$\delta$ are the Tsallis-Cirto entropy parameters. Following
\cite{sheykhi2018modified}, the parameter $\delta$ is taken as
\begin{equation}\label{t2}
\delta=\frac{2-\zeta}{4\,\zeta\,l_P^2}\left(4\,\pi\right)^{1-\zeta},
\end{equation}
where the parameter $\zeta$ satisfies the constraint $0<\zeta<2$.
Using Eqs. \eqref{mh1a}, \eqref{t1}, and \eqref{t2}, together with
the condition $0<\zeta<2$, the mass-to-horizon entropy reduces to
the Tsallis-Cirto entropy for
\begin{equation}\label{t3}
\begin{aligned}
&\eta=2\,\zeta-1,\\
&\lambda=\frac{2-\zeta}{4^\zeta\,\left(2\,\zeta-1\right)}.
\end{aligned}
\end{equation}
The corresponding parameter range, $-1<\eta<3$, follows directly from the condition $0<\zeta<2$. The Barrow entropy is
\cite{barrow2020area}
\begin{equation}\label{b1}
S_B=k_B\,\left(\frac{\pi\,r_s^2}{l_P^2}\right)^{1+\frac{\Xi}{2}},\qquad 0\le\Xi\le1.
\end{equation}
The Barrow entropy reduces to the Bekenstein-Hawking entropy in
the limit $ \Xi=0 $. Comparing Eqs. \eqref{mh1a} and \eqref{b1},
the mass-to-horizon entropy reduces to the Barrow entropy for the
following parameter values
\begin{equation}\label{b2}
\begin{aligned}
&\eta=1+\Xi,\\
&\lambda=\left(\frac{\pi}{l_P^2}\right)^{\frac{\Xi}{2}}\,\frac{\Xi+2}{2\,\left(\Xi+1\right)}.
\end{aligned}
\end{equation}
The corresponding parameter range is $1\le\eta\le2$, which follows
from the condition $0\le\Xi\le1$. Substituting Eq. \eqref{as2}
into Eq. \eqref{mh1a}, and adopting $ \hbar=c=1 $, the
mass-to-horizon entropy becomes
\begin{equation}\label{mh2}
S_{MH}=\frac{2^{2+\eta}\,k_B\,\pi\,\eta\,\lambda}{1+\eta}\,M\,\left(G\,M\right)^\eta.
\end{equation}
Differentiating Eq. \eqref{mh2} with respect to the black hole mass and assuming an infinitesimal mass variation, we obtain
\begin{equation}\label{mh3}
\Delta\,S_{MH}=2^{2+\eta}\,k_B\,\pi\,\eta\,\lambda\,\left(G\,M\right)^\eta\,\Delta\,M,
\end{equation}
We now assume that the entropy variation equals the entropy
associated with the erasure of one bit of information, namely
$k_B\,\ln\,2$. Using the Hawking temperature given by Eq.
\eqref{as4}, the corresponding mass change is
\begin{equation}\label{mh4}
\Delta\,M=\frac{\left(4\,\pi\right)^{\eta-1}}{\eta\,\lambda}\,\left(k_B\,T\right)^\eta\,\ln2.
\end{equation}
For $ \eta=\lambda=1 $, Eq. \eqref{mh4} reduces to
\begin{equation}\label{mh4a}
\Delta\,M=k_B\,T\,\ln\,2,
\end{equation}
which is the result for the Bekenstein-Hawking entropy. For
Landauer's principle to hold, the mass variation must satisfy the
condition $ \Delta\,M\ge k_B\,T\,\ln\,2 $. Eq. \eqref{mh4} then
gives
\begin{equation}\label{mh5}
\left(k_B\,T\right)^{\eta-1}\ge\frac{\eta\,\lambda}{\left(4\,\pi\right)^{\eta-1}}.
\end{equation}
This relation can be analyzed in three cases. For $ \eta>1 $, Eq. \eqref{mh5} gives
\begin{equation}\label{mh6}
T\ge\frac{\left(\eta\,\lambda\right)^{\frac{1}{\eta-1}}}{4\,\pi\,k_B}.
\end{equation}
For Landauer's principle to hold, the Hawking temperature must be
greater than or equal to a critical value. Using the Hawking
temperature given by Eq. \eqref{as4}, the corresponding constraint
on the black hole mass is
\begin{equation}\label{mh7}
M\le\frac{1}{2\,G\,\,\left(\eta\,\lambda\right)^{\frac{1}{\eta-1}}}.
\end{equation}
In this case, Landauer's principle is satisfied only for black
holes with masses smaller than or equal to the critical value
given by Eq. \eqref{mh7}. If $ \eta<1 $, Eq. \eqref{mh5} becomes
\begin{equation}\label{mh8}
T\le\frac{\left(\eta\,\lambda\right)^{\frac{1}{\eta-1}}}{4\,\pi\,k_B}.
\end{equation}
Therefore, the Hawking temperature must be less than or equal to a
critical value for Landauer's principle to remain valid. Using Eq.
\eqref{as4}, the corresponding constraint on the black hole mass
is
\begin{equation}\label{mh9}
M\ge\frac{1}{2\,G\,\,\left(\eta\,\lambda\right)^{\frac{1}{\eta-1}}}.
\end{equation}
This implies that Landauer's principle is satisfied only for black
holes with masses greater than or equal to the critical value
given by Eq. \eqref{mh9}. For $ \eta=1 $, Eq. \eqref{mh5} reduces
to
\begin{equation}\label{mh10}
    \lambda\le1.
\end{equation}
This means, Landauer's principle imposes no constraint on either
the Hawking temperature or the black hole mass. The only
requirement is that the inequality \eqref{mh10} must be satisfied.
Unlike the previous two cases, Landauer's principle constrains
only the parameter $ \lambda $.

We now determine the parameter $ \gamma $ appearing in the area
quantization relation \eqref{bm1} for the mass-to-horizon entropy.
Using the relation $ r_s=\sqrt{A/\left(4\,\pi\right)} $ together
with Eq. \eqref{bm4}, we rewrite the mass-to-horizon entropy
\eqref{mh1} in terms of the horizon area
\begin{equation}\label{mh11} S_{MH}=\frac{\eta\,\lambda\,k_B}{2^\eta\,\left(\eta+1\right)\,\pi^{\frac{\eta-1}{2}}\,l_P^2}\,A^{\frac{\eta+1}{2}}.
\end{equation}
Substituting the discrete area spectrum given by Eq. \eqref{bm1}
into Eq. \eqref{mh11}, the entropy difference between two
consecutive area levels, $ n $ and $ n+1 $, is
\begin{equation}\label{mh12}
\begin{split} \Delta\,S_{MH}=&\frac{k_B\,\eta\,\lambda}{2^\eta\,\left(\eta+1\right)}\left(\frac{l_P}
{\sqrt{\pi}}\right)^{\eta-1}\gamma^{\frac{\eta+1}{2}}\,\Big[\left(n+1\right)^{\frac{\eta+1}{2}} \\
 &-n^{{\frac{\eta+1}{2}}}\Big].
\end{split}
\end{equation}
We assume that the entropy difference given by Eq. \eqref{mh12}
equals the entropy associated with the erasure of one bit of
information, namely $ k_B\,\ln2 $. We then obtain the parameter $
\gamma $
\begin{equation}\label{mh13}
\begin{split} \gamma=&\Big[\frac{2^\eta\,\left(\eta+1\right)\,\ln2}{\eta\,\lambda}\,\left(\frac{\sqrt{\pi}}{l_P}\right)^{\eta-1}\\ &\times\frac{1}{\left(n+1\right)^{\frac{\eta+1}{2}}-n^{\frac{\eta+1}{2}}}\Big]^{\frac{2}{\eta+1}}.
\end{split}
\end{equation}
Using Eqs. \eqref{bm1} and \eqref{mh13}, the relative area
spectrum for the mass-to-horizon entropy is
\begin{equation}\label{mh14}
\begin{split} \frac{\Delta\,A_n}{A_n}=&\frac{A_{n+1}-A_n}{A_n}=\frac{n+1}{n}\\
&\times\left[\frac{\left(n+1\right)^{\frac{\eta+1}{2}}-n^{\frac{\eta+1}{2}}}{\left(n+2\right)^{\frac{\eta+1}{2}}-\left(n+1\right)^{\frac{\eta+1}{2}}}\right]^{\frac{2}{\eta+1}}-1.
\end{split}
\end{equation}
In the limit $ n\gg1 $, Eq. \eqref{mh14} can be approximated by
\begin{equation}\label{mh15} \frac{\Delta\,A_n}{A_n}\simeq\frac{2}{n\,\left(\eta+1\right)}.\end{equation}
This expression shows that the relative spacing between
consecutive area levels decreases as the number $ n $ increases.
Increasing the parameter $ \eta $ further reduces the relative
spacing between adjacent levels. As a result, the area spectrum
approaches the continuous limit more rapidly. In the classical
limit
\begin{equation}\label{mh16} \lim_{n\rightarrow\infty}\,\frac{\Delta\,A_n}{A_n}=0,
\end{equation}
the relative spacing between neighboring area levels vanishes. The
discrete area spectrum therefore approaches a continuous one. In
the next subsection, we apply the same framework to the
Bekenstein-Hawking entropy with correction terms and investigate
the implications of Landauer's principle.
\subsection{Bekenstein-Hawking Entropy with Correction Terms}
One of the most important classes of generalized entropy
formalisms is the corrected Bekenstein-Hawking entropy. These
entropy models incorporate quantum and thermodynamic corrections
into the original Bekenstein-Hawking entropy-area relation
\cite{zhang2008black, banerjee2008quantum1, banerjee2008quantum,
banerjee2008noncommutative, sheykhi2010thermodynamics,
sheykhi2011power}. In these models, the Bekenstein-Hawking entropy
remains the leading contribution, while one or more correction
terms are added as functions of the horizon area. Since many
corrected entropy models share this structure, they can be written
in the general form \cite{sheykhi2010thermodynamics}
\begin{equation}\label{cbh1}
S_{CBH}=S_{BH}+\alpha\,f(A),
\end{equation}
where $ \alpha $ is the correction parameter and $ f(A) $ is a
function of the horizon area. For the quantum corrected entropy, $
f(A) $ is \cite{sheykhi2010thermodynamics}
\begin{equation}\label{cbh2}
f(A)=-\ln\,S_{BH}+\frac{\beta}{\alpha\,S_{BH}},
\end{equation}
where $ \beta $ is a positive constant. For the power law corrected entropy, $ f(A) $ is
\cite{sheykhi2011power}
\begin{equation}\label{cbh3}
f(A)=-K_\alpha\,A^{1-\frac{\alpha}{2}}\,S_{BH},
\end{equation}
where
\begin{equation}\label{cbh4}
K_\alpha=\frac{\left(4\,\pi\right)^{\frac{\alpha}{2}-1}}{\left(4-\alpha\right)\,r_c^{2-\alpha}},
\end{equation}
with $ r_c $ denoting the crossover scale. According to Eqs.
\eqref{as1} and \eqref{as2}, the horizon area $ A $ depends only
on the black hole mass and constant coefficients. Using these
relations together with Eq. \eqref{bm4}, Eq. \eqref{cbh1} becomes
\begin{equation}\label{cbh5}
S_{CBH}=\frac{4\,\pi\,k_B\,\left(G\,M\right)^2}{l_P^2}+\bar{\alpha}\,f(M),
\end{equation}
where $ \bar{\alpha} $ denotes the product of $ \alpha $ and the
corresponding constant coefficients. Assuming an infinitesimal
mass variation and adopting $ \hbar=c=1 $, differentiating Eq.
\eqref{cbh5} with respect to the black hole mass gives
\begin{equation}\label{cbh6}
\Delta\,S_{CBH}=\left[8\,\pi\,G\,k_B\,M+\bar{\alpha}\,f^\prime(M)\right]\,\Delta\,M,
\end{equation}
We assume that the entropy variation given by Eq. \eqref{cbh6}
equals the entropy associated with the erasure of one bit of
information. The corresponding mass variation is
\begin{equation}\label{cbh7}
\Delta\,M=\frac{k_B\,T\,\ln2}{1+\bar{\alpha}\,T\,f^\prime(M)}.
\end{equation}
For $ \bar{\alpha}=0 $, the mass variation reduces to the result
for the Bekenstein-Hawking entropy, which exactly saturates the
Landauer bound. For Landauer's principle to hold, the energy
variation must satisfy the condition $ \Delta\,M\ge k_B\,T\,\ln2
$. Eq. \eqref{cbh7} then gives
\begin{equation}\label{cbh8}
\frac{1}{1+\bar{\alpha}\,T\,f^\prime(M)}\ge1.
\end{equation}
This inequality implies
\begin{equation}\label{cbh9}
-1<\bar{\alpha}\,T\,f^\prime\left(M\right)\le0.
\end{equation}
Since the Hawking temperature is always positive, Eq. \eqref{cbh9} requires
\begin{equation}\label{cbh10}
\bar{\alpha}\,f^\prime\left(M\right)\le0.
\end{equation}
Using Eqs. \eqref{cbh9} and \eqref{cbh10}, we obtain
\begin{equation}\label{cbh11}
T<\frac{1}{\left|\bar{\alpha}\,f^\prime\left(M\right)\right|}.
\end{equation}
The right hand side of Eq. \eqref{cbh11} diverges in the limit $
\alpha\rightarrow0 $. Therefore, this constraint disappears for
the Bekenstein-Hawking entropy. For Landauer's principle to remain
valid, the Hawking temperature must be smaller than a critical
value. Using the mass-temperature relation given by Eq.
\eqref{as4} together with Eq. \eqref{cbh11}, we obtain
\begin{equation}\label{cbh12}
M>\frac{\left|\bar{\alpha}\,f^\prime\left(M\right)\right|}{8\,\pi\,G\,k_B}.
\end{equation}
Hence, the black hole mass must satisfy the condition given by Eq.
\eqref{cbh12}. Only black holes satisfying this condition are
compatible with Landauer's principle within the framework of the
corrected Bekenstein-Hawking entropies.

We now determine the area quantization parameter $ \gamma $.
Because of the general form of Eq. \eqref{cbh1}, an analytical
expression for $ \gamma $ cannot be obtained. The parameter
appears through the horizon area, and its explicit form depends on
the choice of $ f(A) $. We therefore determine $ \gamma $
separately for the quantum corrected entropy and the power law
corrected entropy. Substituting Eq. \eqref{cbh2} into Eq.
\eqref{cbh1}, the quantum corrected entropy becomes
\begin{equation}\label{cq1}
S_{CQ}=S_{BH}-\alpha\,\ln\,S_{BH}+\frac{\beta}{S_{BH}},
\end{equation}
Using Eqs. \eqref{bm1} and \eqref{bm4}, the entropy difference between two consecutive area levels is
\begin{equation}\label{cq2}
\Delta\,S_{CQ}=\frac{k_B\,\gamma}{4}-\frac{4\,\beta}{k_B\,\gamma\,n\,\left(n+1\right)}+\alpha\,\ln\,\left(\frac{n}{n+1}\right).
\end{equation}
We now assume, according to Landauer's principle, that this
entropy variation is equal to $ k_B\,\ln2 $. The area quantization
parameter $ \gamma $ is then obtained as
\begin{equation}\label{cq3}
\begin{split}
\gamma=&2\,\ln2+\frac{2\,\alpha}{k_B}\,\ln\,\left(\frac{n+1}{n}\right)\\
&+\frac{2}{k_B}\,\sqrt{\frac{4\,\beta}{n\,\left(n+1\right)}+\left[k_B\,\ln2+\alpha\,\ln\,\left(\frac{n+1}{n}\right)\right]^2}.
\end{split}
\end{equation}
The positive root is chosen because the parameter $ \gamma $, and
consequently the horizon area, must be positive. Quantum
corrections make the parameter $ \gamma $ depend on the number $ n
$. Using Eq. \eqref{cq3}, the relative area spectrum in the limit
$ n\gg1 $ is
\begin{equation}\label{cq4}
\frac{\Delta\,A_n}{A_n}\simeq\frac{1}{n}.
\end{equation}
In the classical limit $ n\rightarrow\infty $, we have
\begin{equation}\label{cq5}
\lim_{n\rightarrow\infty}\frac{\Delta\,A_n}{A_n}=0.
\end{equation}
Therefore, the relative spacing between adjacent area levels
vanishes in the classical limit, and the area spectrum approaches
a continuous one.

We now repeat the same analysis for the power law corrected
entropy. The only difference is that the entropy difference is
expanded for small values of $ \alpha $ in order to obtain a
closed form expression for $ \gamma $. The Bekenstein-Hawking
entropy is recovered in the limit $ \alpha=0 $. The parameter $
\gamma $ is then obtained as
\begin{equation}\label{plc1}
\begin{split}
\gamma=&\frac{8}{\alpha\,l_P^2\,\left(2\,n+1\right)}\,\Bigg[\pi\,r_c^2\\
&+r_c\,\sqrt{\pi^2\,r_c^2-\alpha\,\pi\,l_P^2\,\left(2\,n+1\right)\,\ln2}\Bigg].
\end{split}
\end{equation}
For the parameter $ \gamma $ to be real, the expression under the
square root must be non-negative. Using Eq. \eqref{plc1}, the
relative area spectrum in the limit $ n\gg1 $ is
\begin{equation}\label{plc2}
\frac{\Delta\,A_n}{A_n}\simeq\frac{2}{n}.
\end{equation}
The classical limit is
\begin{equation}\label{plc3}
\lim_{n\rightarrow\infty}\frac{\Delta\,A_n}{A_n}=0.
\end{equation}
Therefore, the relative spacing between adjacent area levels also
vanishes in the classical limit, and the area spectrum approaches
a continuous one. In the next subsection, we apply Landauer's
principle to other classes of generalized entropy formalisms.
\section{Generalized Entropies with Parameter Constraints Imposed by Landauer's Principle}\label{sec4}
In this section, we apply Landauer's principle to several
generalized entropy formalisms for which the principle imposes
direct constraints on the corresponding entropy parameters.
\subsection{R\'enyi entropy}
R\'enyi entropy is one of the best known generalizations of the
Bekenstein-Hawking entropy and has found widespread applications
in black hole thermodynamics and cosmology in recent years. It is
based on the idea that the long range nature of gravitational
interactions and the non-additive behavior of gravitational
systems may lead to deviations from the standard entropy-area
relation. As a result, R\'enyi entropy provides a suitable
framework for incorporating non-additive effects into
gravitational thermodynamics. It is defined as
\cite{czinner2016renyi}
\begin{equation}\label{r1}
S_R=\frac{k_B}{\nu}\ln\left(1+\nu\,\frac{S_{BH}}{k_B}\right),
\end{equation}
where $ \nu $ is the non-additivity parameter. In the limit $
\nu\rightarrow0 $, this expression reduces to the
Bekenstein-Hawking entropy. Following the same procedure adopted
in the previous sections, the corresponding mass variation is
obtained as ($ \hbar=c=1 $)
\begin{equation}\label{r2}
\Delta\,M=\left(1+\frac{\nu}{16\,\pi\,G\,k_B^2\,T^2}\right)\,k_B\,T\,\ln2.
\end{equation}
For $ \nu=0 $, $ \Delta\,M $ reduces to $ k_B\,T\,\ln2 $, which
coincides with the mass variation for the Bekenstein-Hawking
entropy. In order for Landauer's principle to remain valid, the
condition $ \Delta\,M\ge k_B\,T\,\ln2 $ must hold. Using Eq.
\eqref{r2}, this condition gives
\begin{equation}\label{r3}
\nu\ge0.
\end{equation}
Therefore, Landauer's principle imposes only the constraint $
\nu\ge0 $ on the R\'enyi entropy parameter. It is worth mentioning
that the result obtained here differs from that reported in Ref.
\cite{ananias2026generalized}. In the present work, Landauer's
principle is first implemented through the condition $
\Delta\,M\ge k_B\,T\,\ln2 $. Using Eq. \eqref{r2}, this
requirement directly leads to the constraint $ \nu\ge0 $. Once the
physically allowed parameter range has been established, the
entropy change associated with the erasure of one bit of
information is calculated. It is then used to derive the area
quantization spectrum. In contrast, Ref.
\cite{ananias2026generalized} first imposes the condition $
\Delta\,S_R=k_B\,\ln2 $. This procedure ultimately identifies the
branch $ \nu<0 $ as the physically admissible one. Consequently,
the two approaches differ not only in the order in which
Landauer's principle is implemented, but also in the resulting
allowed domain of the R\'enyi parameter. While the present
analysis admits non-negative values of $ \nu $, Ref.
\cite{ananias2026generalized} selects the negative branch. This
comparison indicates that the physically allowed range of the
R\'enyi parameter should first be determined from the fundamental
Landauer condition $ \Delta\,M\ge k_B\,T\,\ln2 $. Only then should
the condition $ \Delta\,S_R=k_B\,\ln2 $ be imposed. Otherwise, one
may obtain a parameter branch that is incompatible with the
original energy requirement of Landauer's principle.

Using Eqs. \eqref{bm1} and \eqref{r1}, we calculate the entropy
difference between two successive area levels and equate it to the
entropy change associated with the erasure of one bit of
information. The area quantization parameter is
\begin{equation}\label{r4}
\gamma=\frac{4\,\left(1-2^\nu\right)}{\nu\,\left[n\,\left(2^\nu-1\right)-1\right]}.
\end{equation}
Since the area quantization parameter must always be positive, the
condition $ \gamma>0 $ must be satisfied. For $ \nu\ge0 $, the
numerator of Eq. \eqref{r4} is always non-positive. Therefore, the
denominator must also be negative. This requirement gives
\begin{equation}\label{r5}
n<\frac{1}{2^\nu-1}.
\end{equation}
Since $ n\ge1 $, this inequality requires
\begin{equation}\label{r6}
0\le\nu<1.
\end{equation}
Therefore, the condition $ \nu\ge0 $, obtained from Landauer's
principle alone, is not sufficient. Requiring the area
quantization parameter to remain positive further restricts the
R\'enyi parameter to the interval $ 0\le\nu<1 $. Within this
range, the positivity condition for $ \gamma $ becomes $
n<1/\left(2^\nu-1\right) $. Since $ n $ takes only positive
integer values, the allowed quantum levels are
\begin{equation}\label{r7}
n=1,\,2,\,\dots,\,\left\lceil\frac{1}{2^\nu-1}\right\rceil-1,
\end{equation}
where $\left\lceil1/\left(2^\nu-1\right)\right\rceil$ denotes the
ceiling function, namely the smallest integer greater than or
equal to $\frac{1}{2^\nu-1}$. Hence, the ground state ($ n=1 $) is
always allowed, whereas for each fixed value of $ \nu $, only a
finite number of quantum levels satisfy the positivity condition
of $ \gamma $. Moreover, as $ \nu $ increases, the quantity $
1/\left(2^\nu-1\right) $ decreases. Therefore, the number of
allowed quantum levels also decreases. The non-additivity
parameter of the R\'enyi entropy not only modifies the entropy
itself but also directly determines the number of admissible
quantum levels of the event horizon. In the limit $
\nu\rightarrow0 $, one finds
\begin{equation}\label{r8}
\lim_{\nu\rightarrow0}\frac{1}{2^\nu-1}=\infty.
\end{equation}
Therefore, all quantum levels become allowed in this limit. As $
\nu $ approaches unity, only the lowest lying levels remain
admissible. Finally, using Eqs. \eqref{bm1} and \eqref{r4}, the
area spectrum is
\begin{equation}\label{r9}
\frac{\Delta\,A_n}{A_n}=\frac{1}{n\,\left(n+2\right)-2^\nu\,n\,\left(n+1\right)}.
\end{equation}
In the classical limit $ n\rightarrow\infty $, we obtain
\begin{equation}\label{r10}
\lim_{n\rightarrow\infty}\frac{\Delta\,A_n}{A_n}=0.
\end{equation}
Therefore, the relative spacing between adjacent area levels
vanishes in the classical limit, and the area spectrum approaches
a continuous one.
\subsection{Sharma-Mittal entropy}
Sharma-Mittal entropy is a two parameter generalization of the
Bekenstein-Hawking entropy. It combines the main features of the
R\'enyi and Tsallis entropies. Owing to its two independent
parameters, it provides greater flexibility for describing
non-additive systems with long range interactions. For this
reason, it has attracted considerable attention in black hole
thermodynamics and cosmology. The Sharma-Mittal entropy is defined
by \cite{jahromi2018generalized}
\begin{equation}\label{sm1}
S_{SM}=\frac{k_B}{\varrho}\,\left[\left(1+\vartheta\, \frac{S_{BH}}{k_B}\right)^{\frac{\varrho}{\vartheta}}-1\right],
\end{equation}
where $ \varrho $ and $ \vartheta $ are the free entropy
parameters. In the limit $ \varrho\rightarrow\vartheta $, this
expression reduces to the Bekenstein-Hawking entropy. In the limit
$ \varrho\rightarrow0 $, it reduces to the R\'enyi entropy.
Following the same procedure as in the previous sections, the mass
variation associated with Hawking evaporation is obtained as ($
\hbar=c=1 $)
\begin{equation}\label{sm2}
\Delta\,M=\left(1+\frac{\vartheta}{16\,\pi\,G\,k_B^2\,T^2}\right)^{1-\frac{\varrho}{\vartheta}}\,k_B\,T\,\ln2,
\end{equation}
which reduces to the corresponding result for the
Bekenstein-Hawking entropy when $ \varrho=\vartheta $. In this
limit, the Landauer bound is saturated. In order for Landauer's
principle to hold, the mass variation must satisfy $ \Delta\,M\ge
k_B\,T\,\ln2 $. Therefore, Eq. \eqref{sm2} implies
\begin{equation}\label{sm3}
\left(1+\frac{\vartheta}{16\,\pi\,G\,k_B^2\,T^2}\right)^{1-\frac{\varrho}{\vartheta}}\ge1.
\end{equation}
Before analyzing Eq. \eqref{sm3}, the base of the power must remain positive so that the expression is real. This requirement gives
\begin{equation}\label{sm4}
\vartheta>-16\,\pi\,G\,k_B^2\,T^2.
\end{equation}
Under this condition, the monotonicity of the power function shows that Eq. \eqref{sm3} is satisfied only if
\begin{equation}\label{sm5}
\vartheta\ge \varrho.
\end{equation}
Therefore, Landauer's principle imposes two conditions on the
Sharma-Mittal entropy. The first is the reality condition given by
Eq. \eqref{sm4}. The second is the parameter constraint $
\vartheta\ge \varrho $.

We now determine the area quantization parameter associated with
the Sharma-Mittal entropy using the same procedure as in the
previous sections. To obtain an analytic expression for $ \gamma
$, we consider the regime in which the Sharma-Mittal entropy
differs only slightly from the Bekenstein-Hawking entropy.
Specifically, we assume $ \varrho/\vartheta=1+\epsilon $, where $
\epsilon $ is a small parameter. We also assume that $ \vartheta $
is small. Retaining the leading order terms under these
assumptions gives
\begin{equation}\label{sm6}
\begin{split}
\gamma=&\frac{4}{\epsilon\,\vartheta\,k_B\,\left(2\,n+1\right)}\,\Bigg[\frac{\epsilon}{\vartheta}-\epsilon-1\\
&+\sqrt{\left(\epsilon+1-\frac{\epsilon}{\vartheta}\right)^2+2\,\epsilon\,\vartheta\,k_B\left(2\,n+1\right)\,\ln2}\Bigg].
\end{split}
\end{equation}
In the limit $ \epsilon\rightarrow0 $, Eq. \eqref{sm6} reduces to
$ 4\,\ln2 $, which is the area quantization constant of the
Bekenstein-Hawking entropy. Using Eqs. \eqref{bm1} and
\eqref{sm6}, the area spectrum in the limit $ n\gg1 $ is obtained
as
\begin{equation}\label{sm7}
\frac{\Delta\,A_n}{A_n}\simeq\frac{2\,\vartheta\,\sqrt{\epsilon\,\vartheta\,k_B\,n\,\ln2}
+\epsilon\,\left(\vartheta-1\right)+\vartheta}{4\,\vartheta\,n\,\sqrt{\epsilon\,\vartheta\,k_B\,n\,\ln2}}.
\end{equation}
In the classical limit $ n\rightarrow\infty $, one obtains
\begin{equation}\label{sm8}
\lim_{n\rightarrow\infty}\,\frac{\Delta\,A_n}{A_n}=0.
\end{equation}
Therefore, the relative spacing between consecutive area levels
vanishes in the classical limit. As a result, the discrete area
spectrum approaches a continuous one.
\subsection{Loop quantum gravity entropy}
Loop quantum gravity is one of the leading approaches to the
quantization of spacetime geometry. In this framework, geometric
quantities such as the horizon area have discrete spectra.
Counting the quantum microstates associated with a black hole
horizon leads to corrections to the Bekenstein-Hawking entropy.
These corrections may produce deviations from the area law at the
quantum scale. Therefore, this entropy provides an important
framework for investigating black hole thermodynamics. One form of
this entropy, obtained by combining results from non-extensive
statistics and loop quantum gravity, is given by
\cite{ashtekar2000, majhi2017non, liu2022non}
\begin{equation}\label{lqg1}
S_{LQG}=\frac{k_B}{1-q}\left[e^{\left(1-q\right)\,\frac{S_{BH}}{k_B}}-1\right],
\end{equation}
where $ q $ is the non-additivity parameter. In the limit $
q\rightarrow1 $, this expression reduces to the Bekenstein-Hawking
entropy. Following the same procedure as in the previous sections,
the corresponding mass variation is obtained as ($ \hbar=c=1 $)
\begin{equation}\label{lqg2}
\Delta\,M=e^{\frac{q-1}{16\,\pi\,G\,k_B^2\,T^2}}\,k_B\,T\,\ln2.
\end{equation}
For $ q=1 $, Eq. \eqref{lqg2} reduces to $ k_B\,T\,\ln2 $, which
is the corresponding result for the Bekenstein-Hawking entropy. In
order for Landauer's principle to hold, the mass variation must
satisfy $ \Delta\,M\ge k_B\,T\,\ln2 $. Therefore, Eq. \eqref{lqg2}
implies
\begin{equation}\label{lqg3}
e^{\frac{q-1}{16\,\pi\,G\,k_B^2\,T^2}}\ge1.
\end{equation}
Since the exponential function is monotonically increasing, Eq. \eqref{lqg3} is satisfied only if
\begin{equation}\label{lqg4}
q\ge1.
\end{equation}
Therefore, Landauer's principle imposes the lower bound $ q\ge1 $
on the non-additivity parameter. We now determine the
corresponding area quantization parameter. Using Eqs. \eqref{bm1},
\eqref{bm4}, and \eqref{lqg1}, we first calculate the entropy
difference between two successive area levels and equate it to the
entropy change associated with the erasure of one bit of
information. To obtain an analytic expression for $ \gamma $, we
further assume that $ q=1+\bar{\epsilon} $, where $ \bar{\epsilon}
$ is a small parameter. This assumption corresponds to the regime
in which the entropy differs only slightly from the
Bekenstein-Hawking entropy. Under these assumptions, the area
quantization parameter is obtained as
\begin{equation}\label{lqg5}
\gamma=\frac{4}{\bar{\epsilon}\,k_B\,\left(2\,n+1\right)}\,\left[1-\sqrt{1-2\,\bar{\epsilon}\,k_B\,\ln2\,\left(2\,n+1\right)}\right].
\end{equation}
In the limit $ \bar{\epsilon}\rightarrow0 $, Eq. \eqref{lqg5} reduces to $ 4\,\ln2 $, which is the area quantization constant of the Bekenstein-Hawking entropy.

Using Eqs. \eqref{bm1} and \eqref{lqg5}, the classical limit of the area spectrum is
\begin{equation}\label{lqg6}
\lim_{n\rightarrow\infty}\,\frac{\Delta\,A_n}{A_n}=0.
\end{equation}
Therefore, the relative spacing between consecutive area levels
vanishes in the classical limit. As a result, the discrete area
spectrum approaches a continuous one.
\subsection{Hypergeometric entropy}
Hypergeometric entropy is one of the generalized forms of the
Bekenstein-Hawking entropy. It was proposed to incorporate
corrections arising from non-Newtonian behavior at large scales
and from the interplay between gravity and thermodynamics. Within
this framework, modifying the entropy-area relation leads to a
general expression for the horizon entropy that includes several
well known entropy formalisms as special cases. In particular, it
reduces to the R\'enyi entropy \cite{czinner2016renyi} and to the
dual form of the Kaniadakis entropy \cite{abreu2021black} for
suitable choices of the free parameters. The hypergeometric
entropy is defined as \cite{sheykhi2025mond}
\begin{equation}\label{h1}
S_H=S_{BH}\,{}_2F_1\left(\frac{1}{\sigma},\,\frac{1}{\sigma},\,\frac{\sigma+1}{\sigma},\,-\left(\frac{\xi\,S_{BH}}{k_B}\right)^\sigma\right),
\end{equation}
where $ {}_2F_1 $ denotes the Gauss hypergeometric function,
$\sigma $ is a constant parameter, and $ \xi $ is a parameter
associated with the transition scale. In the limit $ \sigma=1 $,
this expression reduces to the R\'enyi entropy, whereas for $
\sigma=2 $, it reduces to the dual form of the Kaniadakis entropy.
In the limit $ \xi\rightarrow0 $, it recovers the
Bekenstein-Hawking entropy.

Following the same procedure adopted in the previous sections, the corresponding mass variation is obtained as ($ \hbar=c=1 $)
\begin{equation}\label{h2}
\Delta\,M=\left[1+\left(\frac{\xi}{16\,\pi\,G\,k_B^2\,T^2}\right)^\sigma\right]^{\frac{1}{\sigma}}\,k_B\,T\,\ln2.
\end{equation}
In the limit $ \xi\rightarrow0 $, this expression reduces to $ k_B\,T\,\ln2 $. Since Landauer's principle requires $ \Delta\,M\ge k_B\,T\,\ln2 $, one obtains
\begin{equation}\label{h3}
\left[1+\left(\frac{\xi}{16\,\pi\,G\,k_B^2\,T^2}\right)^\sigma\right]^{\frac{1}{\sigma}}\ge1.
\end{equation}
This inequality is satisfied only if the parameter $ \sigma $ is positive. In addition, the parameter $ \xi $ must satisfy
\begin{equation}\label{h4}
\xi\ge0.
\end{equation}
so that the expression remains real. Therefore, applying
Landauer's principle to the hypergeometric entropy imposes the
parameter constraints
\begin{equation}\label{h5}
\sigma>0,\qquad \xi\ge0.
\end{equation}
We now determine the area quantization parameter. To this end, we
first expand the hypergeometric function in the regime $
\left[\left(\xi\,A\right)/\left(4\,l_P^2\right)\right]^\sigma\ll1
$. We then evaluate the entropy difference between two consecutive
levels, $ n $ and $ n+1 $
\begin{equation}\label{h6}
\begin{split}
S_{n+1}-S_n=&\frac{1}{4}\,k_B\,\gamma\,\Bigg[1+\frac{\left(\gamma\,\xi\right)^\sigma}{4^\sigma\,\sigma\,\left(\sigma+1\right)}\,\Big(n^{\sigma+1}\\
&-\left(n+1\right)^{\sigma+1}\Big)\Bigg].
\end{split}
\end{equation}
To obtain an analytic expression for the area quantization
parameter $ \gamma $, we set $ \sigma=1+\tilde{\epsilon} $, where
$ |\tilde{\epsilon}|\ll1 $. We then expand Eq. \eqref{h6} to first
order in both $ \xi $ and $ \tilde{\epsilon} $, while neglecting
terms of order $ \mathcal{O}(\xi\tilde{\epsilon}) $. Equating the
resulting entropy difference to the entropy associated with the
erasure of one bit of information gives
\begin{equation}\label{h7}
\gamma=\frac{4}{\xi\,\left(2\,n+1\right)}\,\left[1-\sqrt{1-2\,\xi\,\ln2\,\left(2\,n+1\right)}\right].
\end{equation}
In the limit $ \xi\rightarrow0 $, the area quantization parameter
reduces to $ \gamma=4\,\ln2 $, which is the area quantization
constant of the Bekenstein-Hawking entropy. Using Eqs. \eqref{bm1}
and \eqref{h7}, the area spectrum in the regime $ n\gg1 $ is
obtained as
\begin{equation}\label{h8}
\frac{\Delta\,A_n}{A_n}\simeq\frac{1}{2\,n}.
\end{equation}
In the classical limit $ n\rightarrow\infty $, one obtains
\begin{equation}\label{h9}
\lim_{n\rightarrow\infty}\frac{\Delta\,A_n}{A_n}=0.
\end{equation}
Therefore, in the classical limit, the relative spacing between
consecutive area levels vanishes, and the area spectrum approaches
a continuous one.
\section{Entropies Incompatible with Landauer's Principle}\label{sec5}
Kaniadakis entropy is a generalized entropy formalism derived from
the $\kappa$-deformed statistical framework introduced by
Kaniadakis. This framework is inspired by relativistic kinematics
and provides a deformation of Boltzmann-Gibbs statistics while
recovering the standard theory in the limit $\kappa\rightarrow0$.
Owing to the long-range nature of gravitational interactions and
the non-additive properties of gravitational systems, Kaniadakis
entropy has been widely considered as an alternative
generalization of the Bekenstein-Hawking entropy in black hole
thermodynamics. It is defined by \cite{kaniadakis2002statistical,
kaniadakis2006towards}
\begin{equation}\label{k1}
S_K=\frac{k_B}{\kappa}\sinh\left(\kappa\,\frac{S_{BH}}{k_B}\right).
\end{equation}
where $ \kappa $ is the Kaniadakis deformation parameter. In the
limit $ \kappa\rightarrow0 $, the hyperbolic sine function can be
expanded to recover the Bekenstein-Hawking entropy.

Following the same procedure adopted in the previous sections, and
assuming that the mass variation is infinitesimal, the
corresponding mass variation is obtained by using the Kaniadakis
entropy together with the entropy change associated with the
erasure of one bit of information ($ \hbar=c=1 $)
\begin{equation}\label{k2}
\Delta\,M=\frac{k_B\,T\,\ln2}{\cosh\left(\frac{\kappa}{16\,\pi\,G\,k_B^2\,T^2}\right)}.
\end{equation}
For $ \kappa=0 $, this expression reduces to $ k_B\,T\,\ln2 $,
which is the corresponding result for the Bekenstein-Hawking
entropy. Imposing the Landauer condition $ \Delta\,M\ge
k_B\,T\,\ln2 $ yields
\begin{equation}\label{k3}
\frac{1}{\cosh\left(\frac{\kappa}{16\,\pi\,G\,k_B^2\,T^2}\right)}\ge1.
\end{equation}
Since the inequality $ \cosh(x)\ge1 $ holds for every real value
of $ x $, Eq. \eqref{k3} is satisfied only when the equality
holds, namely for $ \kappa=0 $. This limit corresponds to the
Bekenstein-Hawking entropy. Therefore, the standard Kaniadakis
entropy does not satisfy Landauer's principle for any non-zero
value of the deformation parameter $ \kappa $. Among the
generalized entropy formalisms considered in this work, it is
compatible with Landauer's principle only in the limit where it
reduces to the Bekenstein-Hawking entropy.
\section{Comparison of Generalized Entropy Models within the Landauer Framework}\label{sec6}
\begin{figure}[ht]
\centering
\includegraphics[width=1\linewidth]{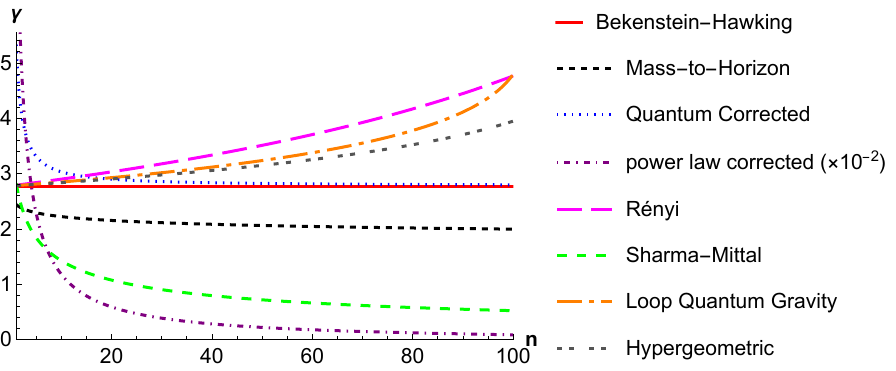}
\caption{Comparison of the area quantization parameter $\gamma$
obtained from different entropy formalisms. The
Bekenstein--Hawking entropy predicts a constant quantization
parameter, $\gamma_{BH}=4\,\ln2$, corresponding to an equally
spaced area spectrum. In contrast, generalized entropy models lead
to an $n$-dependent quantization parameter, resulting in
non-uniform spacing between successive area eigenvalues. For
clarity, the curve corresponding to the Power-law Corrected
entropy is plotted after rescaling by a factor of $10^{-2}$. The
parameter values used in the figure are $l_P=1$, $k_B=1$,
$\eta=1.1$, $\lambda=1.1$, $\alpha=0.6$, $\beta=0.6$, $r_c=5.5$,
$\nu=0.006$, $\epsilon=0.6$, $\theta=0.6$, $\bar{\epsilon}=0.6$,
and $\xi=0.003$.} \label{fig1}
\end{figure}

In this section, the generalized entropy models investigated in
this work are compared within a unified framework based on
Landauer's principle. Tab. \ref{tab:constraints} summarizes the
constraints imposed by Landauer's principle on each entropy model,
while Fig. \ref{fig1} compares the corresponding area quantization
parameters. For completeness, we also recall that, for all entropy
models compatible with Landauer's principle, the area spectrum has
been examined in the large-area-level limit ($n\gg1$). In this
limit, the relative spacing between consecutive area levels is
proportional to $1/n$ and therefore vanishes as
$n\rightarrow\infty$.

Tab. \ref{tab:constraints} summarizes the generalized entropy
models considered in this work together with their entropy
expressions and the corresponding constraints imposed by
Landauer's principle. This comparison facilitates a direct
assessment of the different entropy formalisms and clearly
illustrates how Landauer's principle constrains each generalized
entropy model.
\begin{table*}[htbp]
\centering \caption{Summary of constraints imposed by Landauer's
principle on generalized entropy formalisms. The
Bekenstein-Hawking entropy, which exactly saturates the bound, is
excluded as the reference case. The complete derivations are
provided in the main text.} \label{tab:constraints}
\renewcommand{\arraystretch}{1.3}
\begin{tabular}{|>{\centering\arraybackslash}p{3.5cm}|>{\centering\arraybackslash}p{6.3cm}|>{\centering\arraybackslash}p{7.5cm}|}
\hline \textbf{Entropy Model} & \textbf{Entropy Formula} &
\textbf{Constraint Imposed by Landauer's Principle} \\ \hline
\hline Mass-to-Horizon
&\(S_{MH}=\frac{2\,\eta\,\lambda}{\eta+1}\,r_s^{\eta-1}\,S_{BH}\)
& \begin{tabular}{@{}>{\centering\arraybackslash}c@{}}
\textcolor{blue}{Temperature (and mass) constraint:}\\ If
\(\eta>1\): \(T \ge T_{\text{crit}}\) or \(M\le
M_{\text{crit}}\)\\ If \(\eta<1\): \(T \le T_{\text{crit}}\) or
\(M\ge M_{\text{crit}}\)\\ If \(\eta=1\): \(\lambda \le 1\) \\
where \(T_{\text{crit}} =
\dfrac{(\eta\,\lambda)^{\frac{1}{\eta-1}}}{4\,\pi\,k_B}\) and
\(M_{\text{crit}}
=\frac{1}{2\,G\,\,\left(\eta\,\lambda\right)^{\frac{1}{\eta-1}}}\)
\end{tabular} \\ \hline Bekenstein-Hawking with Correction Terms &
\( S_{CBH} = S_{BH} + \alpha\,f(A) \) &
\begin{tabular}{@{}>{\centering\arraybackslash}c@{}}
\textcolor{blue}{Temperature (and mass) constraint:}\\ \( T <
\dfrac{1}{|\bar{\alpha}\,f'(M)|} \) \\ or equivalently: \\ \( M >
\dfrac{|\bar{\alpha}\,f'(M)|}{8\,\pi\,G\,k_B} \) \end{tabular} \\
\hline Rényi & \( S_{R} = \frac{k_B}{\nu}\,\ln\,\left(1+\nu\,
\frac{S_{BH}}{k_B}\right) \) &
\begin{tabular}{@{}>{\centering\arraybackslash}c@{}}
\textcolor{blue}{Parameter constraint:}\\ \( \nu \ge 0 \) \\ (with
\(\gamma>0\) further requiring \(0 \le \nu < 1\)) \end{tabular} \\
\hline Sharma-Mittal & \( S_{SM} =
\frac{k_B}{\varrho}\,\left[\left(1+\vartheta\,
\frac{S_{BH}}{k_B}\right)^{\frac{\varrho}{\vartheta}} - 1\right]
\) & \begin{tabular}{@{}>{\centering\arraybackslash}c@{}}
\textcolor{blue}{Parameter constraint:}\\ \( \vartheta \ge \varrho
\) \\ and \( \vartheta > -16\,\pi\,G\,k_B^2\,T^2 \) \end{tabular}
\\ \hline Loop Quantum Gravity & \( S_{LQG} =
\frac{k_B}{1-q}\,\left[e^{(1-q)\,\frac{S_{BH}}{k_B}} - 1\right] \)
& \begin{tabular}{@{}>{\centering\arraybackslash}c@{}}
\textcolor{blue}{Parameter constraint:}\\ \( q \ge 1 \)
\end{tabular} \\ \hline Hypergeometric
&\(S_H=S_{BH}\,{}_2F_1\left(\frac{1}{\sigma},\,\frac{1}{\sigma},\,\frac{\sigma+1}{\sigma},\,-\left(\frac{\xi\,S_{BH}}{k_B}\right)^\sigma\right)\)
& \begin{tabular}{@{}>{\centering\arraybackslash}c@{}}
\textcolor{blue}{Parameter constraint:}\\ \( \sigma > 0 \) \quad
and \quad \( \xi \ge 0 \) \end{tabular} \\ \hline Kaniadakis & \(
S_{K} = \frac{k_B}{\kappa} \sinh\,\left(\kappa
\frac{S_{BH}}{k_B}\right) \) &
\begin{tabular}{@{}>{\centering\arraybackslash}c@{}}
\textcolor{blue}{Incompatible:}\\ Only recovered in the limit
\(\kappa = 0\),\\ which reduces to the Bekenstein-Hawking entropy.
\end{tabular} \\ \hline \end{tabular} \end{table*}

Fig. \ref{fig1} compares the area quantization parameter predicted
by the different entropy models. For the Bekenstein-Hawking
entropy, the quantization parameter remains constant,
$\gamma_{BH}=4,\ln2$, corresponding to an equally spaced area
spectrum that is independent of the area level number $n$. In
contrast, all generalized entropy models predict an $n$-dependent
quantization parameter, implying that the spacing between
successive area levels is generally non-uniform. However, the
dependence of $\gamma$ on $n$ differs from one entropy model to
another.

The R'enyi entropy, the Loop Quantum Gravity entropy, and the
Hypergeometric entropy exhibit an increasing quantization
parameter as the area level number increases. In contrast, the
Mass-to-Horizon entropy, the Sharma-Mittal entropy, and the
Power-law Corrected entropy exhibit a decreasing behavior,
indicating that the quantization parameter gradually becomes
smaller at higher area levels. The Quantum Corrected entropy
behaves differently from the other generalized entropy models.
After a rapid initial variation, it quickly approaches the
Bekenstein-Hawking value and remains nearly constant over a wide
range of area levels. Therefore, although the Bekenstein-Hawking
entropy predicts a constant area quantization parameter,
generalized entropy models generally lead to an
area-level-dependent quantization parameter.
\section{Conclusions and outlook}\label{sec7}
In this work, we employed Landauer's principle as a fundamental
thermodynamic criterion to investigate the consistency of
generalized black hole entropy formalisms. We first revisited the
Bekenstein-Mukhanov approach, according to which the black hole
horizon area is quantized as $A_n=\gamma\,l_P^2\,n$, with the area
quantization parameter given by $ \gamma=4\ln k $. We then showed
that, if each transition between two consecutive horizon area
levels corresponds to the erasure of one bit of information,
Landauer's principle predicts the area quantization constant $
\gamma=4\ln2 $. Comparing these two results demonstrates that the
Bekenstein-Mukhanov approach is fully consistent with the
Landauer-based framework only for $ k=2 $, in which case both
approaches predict the same spacing between consecutive area
levels. Moreover, under the same assumption, the emitted energy
associated with each Hawking evaporation step was shown to
saturate the Landauer bound for the Bekenstein-Hawking entropy.

The same framework was then applied to a broad class of
generalized entropy formalisms. Depending on the mathematical
structure of the entropy, Landauer's principle gives rise to two
different types of physical constraints. For the Mass-to-Horizon
entropy and the Bekenstein-Hawking entropy with correction terms,
it leads directly to constraints on the Hawking temperature and,
consequently, on the black hole mass. In contrast, for the
R\'enyi, Sharma-Mittal, Loop Quantum Gravity, and Hypergeometric
entropies, the constraints are imposed directly on the free
parameters of the corresponding entropy formalisms. In the case of
the R\'enyi entropy, requiring the area quantization parameter to
remain positive further restricts the allowed parameter range and
shows that the number of admissible quantum levels depends
explicitly on the non-additivity parameter.

The area quantization parameter and the corresponding area
spectrum were also derived for all entropy formalisms compatible
with Landauer's principle. Unlike the Bekenstein-Hawking entropy,
generalized entropy models generally predict an area quantization
parameter that depends on the quantum number. Nevertheless, for
every compatible entropy model considered in this work, the
relative spacing between consecutive area levels tends to zero in
the classical limit, indicating that the discrete area spectrum
correctly approaches a continuous one at macroscopic scales.

Finally, the analysis of the standard Kaniadakis entropy showed
that it does not satisfy the fundamental Landauer condition,
$\Delta M \ge k_B T \ln2$, for any non-zero value of the
deformation parameter. Therefore, among the generalized entropy
formalisms investigated in this work, the standard Kaniadakis
entropy is compatible with Landauer's principle only in the limit
in which it reduces to the Bekenstein-Hawking entropy.

Overall, the results of this work show that Landauer's principle
provides a unified thermodynamic framework for examining
generalized black hole entropy formalisms while establishing a
direct connection between black hole thermodynamics and
information theory. Within this framework, the principle either
constrains the Hawking temperature and the black hole mass or
determines the physically admissible ranges of the entropy
parameters, depending on the mathematical structure of the entropy
under consideration. Future investigations may extend the present
framework to other generalized entropy formalisms, as well as to
rotating and charged black holes, in order to further explore the
role of Landauer's principle in black hole thermodynamics.
\acknowledgments{We thank Shiraz University Research Council.}

\end{document}